\documentclass[aps,prl,reprint,superscriptaddress]{revtex4-2}
\usepackage{graphicx}
\usepackage{times}
\usepackage{physics}
\usepackage{braket}
\usepackage{hyperref}
\usepackage[title]{appendix}
\usepackage{dsfont}
\usepackage{xcolor}
\usepackage{enumitem}
\usepackage{soul}
\usepackage{relsize}
\usepackage{amsfonts}
\usepackage{amssymb}
\graphicspath{{"../figures/"}}

\begin{document}

\title{Manipulating the topological spin of Majoranas}

\author{Stijn R. de Wit\textsuperscript{1,*}}
\author{Emre Duman\textsuperscript{1,2,*}}
\author{A. Mert Bozkurt\textsuperscript{3,4}}
\author{Alexander Brinkman\textsuperscript{1}}
\author{\.Inan\c{c} Adagideli\textsuperscript{1,2,5,$\dagger$}}

\noaffiliation
\affiliation{MESA+ Institute for Nanotechnology, University of Twente, The Netherlands}
\affiliation{Faculty of Engineering and Natural Sciences, Sabanci University, Istanbul, Turkey}
\affiliation{QuTech, Delft University of Technology, Delft 2600 GA, The Netherlands}
\affiliation{Kavli Institute of Nanoscience, Delft University of Technology, P.O. Box 4056, 2600 GA Delft,
The Netherlands}
\affiliation{T\"UB\.ITAK Research Institute for Fundamental Sciences, 41470 Gebze, Turkey
\\ \ \\
\textsuperscript{*}\,These authors contributed equally\\
\textsuperscript{$\dagger$}\,
{Correspondence to: adagideli@sabanciuniv.edu}
\\ \ \\
}
\begin{abstract}
The non-Abelian exchange statistics of Majorana zero modes make them interesting for both technological applications and fundamental research. Unlike their non-Abelian counterpart, the Abelian contribution, $e^{i\theta}$, where $\theta$ is directly related to the Majorana's topological spin, is often neglected. However, the Abelian exchange phase and hence the topological spin can differ from system to system. 
For vortices in topological superconductors, the Abelian exchange phase is interpreted as an Aharonov-Casher phase arising from a vortex encircling a $e/4$ charge.
In this work, we show how this fractional charge, and hence the topological spin, can be manipulated through the control of device geometry, introducing an additional control knob for topological quantum computing. To probe this effect, we propose a vortex interference experiment that reveals the presence of this fractional charge through shifts in the critical current.
\end{abstract}
\maketitle

The non-Abelian exchange statistics of Majorana zero modes (MZMs) is central to their technological promise in the field of fault-tolerant~\cite{kitaev_fault-tolerant_2003} topological quantum computation~\cite{castagnoli_notions_1993, nayak_non-abelian_2008, lahtinen_short_2017, beenakker_search_2020}. This has sparked an ongoing hunt for condensed matter systems that realize MZMs~\cite{beenakker_search_2013,lutchyn_majorana_2018, yazdani2023}. A plethora of candidate platforms exist, ranging from MZMs emerging in strong spin–orbit semiconductor-superconductor heterostructures~\cite{sau_generic_2010, lutchyn_majorana_2010,oreg_helical_2010,mourik_signatures_2012, m_aghaee_interferometric_2025}, fractional quantum Hall $\nu=5/2$ systems~\cite{moore_nonabelions_1991}, ferromagnetic atomic chains~\cite{nadj_perge2014}, chains of quantum dots~\cite{sau_realizing_2012,leijnse_parity_2012,haaf2023Engineering,zatelli_robust_2024, bordin_enhanced_2025}; or MZMs bound to vortices in intrinsic $p$-wave superconductors~\cite{read_paired_2000,ivanov_non-abelian_2001}, planar Josephson junctions~\cite{shabani_two-dimensional_2016,pientka_topological_2017,schiela_progress_2024}, or superconductor-topological insulator proximity systems~\cite{fu_superconducting_2008, cook_majorana_2011}.

Independent of how they are physically realized, the MZMs obey non-trivial anyonic exchange statistics. 
The exchange of two MZMs from different fermionic modes
is described by the unitary operator 
$\mathcal{U} = e^{i\theta}\mathcal{U}_0.$ 
Here, $\mathcal{U}_0$, common to all Majorana platforms, describes the universal non-Abelian exchange~\footnote{The explicit form of this unitary, when it pertains to the exchange of two Majorana modes $\hat\gamma_{1,2}$, reads $\mathcal{U}_0=1+\hat\gamma_1 \hat\gamma_2$, see~\cite{beenakker_search_2020}}.
The Abelian contribution to the exchange, $e^{i\theta}$, has attracted less attention. 
This phase is related to the phase $e^{i2\pi s}$ that arises from the process of ``twisting an anyon around itself"~\cite{simon_topological_2023,nava_non-abelian_2024, ariad_signatures_2017}. Here, $s$ is the topological spin of the anyon. For bosons and fermions, $s$ takes the well known integer and half-integer values, yielding exchange phases of +1 and -1, respectively.
In contrast, anyons such as MZMs bound to vortices can have more exotic fractional topological spin; a well-known example is that of Ising anyons with $\theta = \pi/8$ corresponding to a topological spin of $1/16$. 
Although the exchange phase originating from the topological spin is a fundamental property of MZMs, finding experimentally measurable consequences proved to be a hard problem \cite{nava_2024}. 

\begin{figure}
    \centering
    \includegraphics[width=\columnwidth]{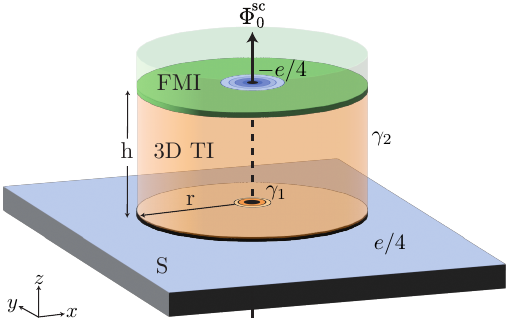}   
    \caption{\textbf{S/3D~TI/FMI proximity system.} By proximity, a magnetic energy gap (an s-wave superconductive pair potential) is introduced in a 3D topological insulator from the top (bottom) by a feromagnetic insulator (s-wave superconductor). A localized $-e/4$ Fermi-sea charge accumulates around the vortex core (carrying $\Phi_0^{\rm SC}=h/2e$ flux) at the top interface and Majorana zero modes appear around the vortex core at the bottom interface  ($\gamma_{1}$) and the outer-surface ($\gamma_{2}$). The magnetic proximity system can, without loss of generality, be replaced by an intrinsic magnetic 3D TI.}
    \label{fig:3D_TI_with_vortex}
\end{figure}

In this manuscript, we show how one can manipulate the topological spin of the MZMs, and we propose an experimental setup to detect it. Among other things, this opens up the way to controllably add fractionally quantized phases to the quantum state of the system.
We focus on MZMs in proximitized topological superconductors bound to (Abrikosov or Josephson) vortices that carry a single superconducting flux quantum, $\Phi_0^{\text{sc}} =h/2e$ (e.g., Fig.~\ref{fig:3D_TI_with_vortex}). 
In these vortex systems, the topological spin is related to the Aharonov-Casher~\cite{aharonov_topological_1984} phase acquired by interference of a vortex around a fractional charge bound to a second vortex~\cite{grosfeld_proposed_2011,grosfeld_observing_2011,ariad_signatures_2017}. Hence, measuring the charge bound to the vortex is equivalent to measuring the topological spin. 

Our analysis proceeds in three stages. First, we mathematically establish the link between the topological spin and the fractional Fermi-sea charge bound to a vortex. 
Using this link, we then reveal fundamental differences in the topological spin across various platforms. 
We show that for a vortex in prototypical 2D models that feature topological superconductivity,
a MZM and a system-dependent charge co-localize. In contrast,  we show that in three-dimensional topological insulator (3D~TI) heterostructures, the charge becomes fractionally quantized to $-e/4$ and this charge and the MZM can be manipulated to localize on opposite surfaces. Finally, we propose a vortex-interference experiment that predicts a universal critical current response, offering a direct probe of this fractional charge.

Our starting point is the mean-field Hamiltonian of a topological material in proximity to a superconductor:
\begin{equation}
\label{Eq:BCS}
   \mathcal{H}(\phi_0)= \sum_\mathbf{k}\psi^\dagger_{\mathbf{k}} 
   H_0({\mathbf{k}})\psi^{\phantom{\dagger}}_{\mathbf{k}} + \sum_{j} 
   \Delta_0 e^{i\varphi_j +i \phi_0}\psi_{j\uparrow}\psi_{j\downarrow} + \text{h.c.}.
\end{equation}
Here $\psi^\dagger_\mathbf{k}$ creates a fermion with momentum $\mathbf{k}$ and $\psi^\dagger_{j,\sigma}$ creates a fermion at the site $j$ with spin $\sigma$, $H_0(\mathbf{k})$ is the first quantized Hamiltonian of the topological material,  
$\Delta_0$ is the induced pair potential, 
and $\varphi_j$ is the phase of the order parameter at site $j$. In the absence of a vortex $\varphi_i=0$, and in the presence of a vortex
$\varphi_i=\arctan\frac{y_i}{x_i}$, for $(x_i,y_i,z_i)$ defined on a lattice along the unit vectors defined in Fig.~\ref{fig:3D_TI_with_vortex}. The gauge field in the presence of a vortex is introduced via the minimal coupling ${\mathbf{k}} \rightarrow {\mathbf{k}}-e{\mathbf{A}}$. 

We now establish the relation between the charge bound to the vortex and the topological spin. Note that, in the case of a vortex at the origin, cycling the phase $\phi_0 \rightarrow \phi_0 + 2\pi$,
circulates an antivortex (at infinity) around the vortex~\footnote{The braiding of the edge MZM around the center MZM is achieved by a 360 degree real space rotation around the $z$-axis. For a rotational symmetric sample, this is equivalent to a $2\pi$ global phase cycle}. 
Then the topological spin is the angle of the (Abelian) Berry phase accumulated during this double-exchange process in units of $2\pi$.  
Changing $\phi_0 $ is established by the
unitary transformation $U(\phi)=\exp(i\phi \hat{N}/2)$, where $\hat{N}=\sum_{i,\sigma}\psi^\dagger_{i,\sigma}\psi^{\phantom{\dagger}}_{i,\sigma}$ is the total number operator of fermions in the topological material. In  order to see this notice that $U$ transforms the Hamiltonian as $H(\phi_0+\phi)= U(\phi) H(\phi_0) U^\dagger(\phi)$. Hence, the BCS ground state $\vert \mathrm{BCS}_{\phi_0} \rangle$ of the Hamiltonian $\mathcal{H}(\phi_0)$ transforms as: 
\begin{equation}\label{eq:BCS}
    \ket{\mathrm{BCS}_{\phi_0+\phi}}=\exp\left(\frac{i\hat N\phi}{2}\right)\ket{\mathrm{BCS}_{\phi_0}}.
\end{equation}
Cycling the phase $\phi_0$ by $2\pi$ takes the system on an adiabatic path along which the Berry connection is
\[
\mathcal{A}(\phi_0)=-i\bra{\mathrm{BCS}_{\phi_0}}\partial_{\phi_0}\ket{\mathrm{BCS}_{\phi_0}} =\frac{1}{2}\langle\hat N\rangle
\]
which leads to the relation between the topological spin and 
the Fermi sea charge ($e\langle \hat N\rangle$):
\begin{equation}\label{eq:berry connection phi0}
    \exp 4i\pi s  = \exp i\int_0^{2\pi}\mathcal{A}(\phi_0) d\phi_0 = \exp i\pi\langle \hat N\rangle.
\end{equation}
We see that adding an integer number of Cooper pairs does not change this phase while adding an electron gives a ${}-{}$ sign. In this work, we are concerned with the fractional part, that is,
\begin{equation}\label{eq: BP-N}
    4s \equiv \langle \hat N\rangle \pmod{2}.
\end{equation}
We explore how well this relation holds numerically in the Supplemental Material~\cite{supp_read}.
In the remainder of this manuscript we focus on calculating the Fermi-sea-charge density for different interfaces of topological materials and superconductors.
The tight-binding simulations~\cite{supp} are performed using Kwant~\cite{Groth_2014} and the adaptive sampling algorithm Adaptive~\cite{Nijholt2019}. We take $e>0$ to be the elementary charge.
\begin{figure}
    \centering
    \includegraphics[width=\columnwidth]{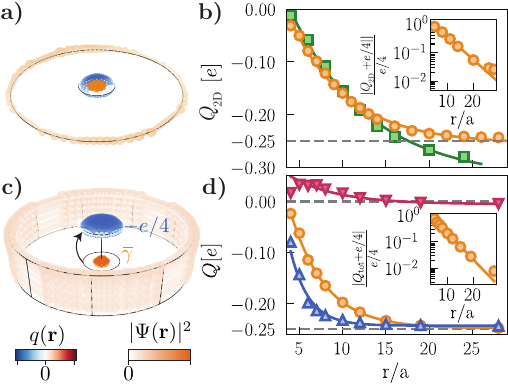}  
    \caption{\textbf{Charge and MZM densities in 2D (top row) and 3D (bottom row).} 
    \textbf{a)} Charge and MZM densities as a result of a $\Phi_0^{\text{sc}}=h/2e$ vortex in a 2D TI system. Colorbars for the charge $q(\mathbf{r})$ and zero mode density $|\psi(\mathbf{r})|^2$ are shared across the figure.
    \textbf{b)} Integrated charge density as a function of system size $r$ in units of the lattice space, $a$, for the two band QAH model (green squares) and four band model being the 2D limit of the 3D~TI (orange circles). Inset: relative deviation of integrated charge from $-e/4$.
    \textbf{c-d)} Similar to \textbf{a-b)} but for a 3D S/3D~TI/FMI heterostructure and with additional markers for the charge integrated on the bottom surface(red triangle) and the top surface(blue triangle).}
    \label{fig:increasing_thickness}
\end{figure}

We first focus on a QAH system in proximity to an s-wave superconductor. For the normal state Hamiltonian, we adopt the prototypical Qi-Hughes-Zhang\cite{qi_chiral_2010} two band model,
\begin{equation}\label{eq:2D_2band_Hamiltonian}
H^{\text{2D}}_0 = \sum_{i=x,y}t \sin k_i \sigma_i + \beta(\boldsymbol{k}) \sigma_z -\mu\sigma_0.
\end{equation}
Here, the first term represents spin-momentum coupling,
the second term, $\beta(\boldsymbol{k})=\beta_0+\beta_2\sum_{i=x,y}(1-\cos k_i)$, is a combination of the magnetization ($\beta_0$) and regularization ($\beta_2$), and $\mu$ is the chemical potential. We place this system in proximity to a conventional s-wave superconductor, resulting in an effective proximity gap $\Delta_0$. The system is in the $N=1$ topological-superconductor phase for $\Delta_0^2+\mu^2>\beta_0^2$~\cite{qi_chiral_2010}. In this phase, a vortex of flux $\Phi_0^{\text{sc}}$ binds a MZM localized at the vortex core and another at the outer edge\cite{read_paired_2000, qi_chiral_2010}.
We now calculate the Fermi sea charge around the vortex and find that a non-zero charge co-localizes with the MZM, see Fig.~\ref{fig:increasing_thickness}~a and the green curve in panel b. However, 
we find that this charge is sensitive to variations in $\beta_0$, $\Delta_0$ and $\mu$ in this QAH model. 

Next, we replace the simple two band model by a more realistic four band model -- which we interpret as the 2D limit of a 3D~TI~\cite{asmar_topological_2018, zhang_crossover_2010,van_veen_observation_2025}. In this case, we recover a fractionally quantized charge  of $-e/4$ (see the orange circles in Fig.~\ref{fig:increasing_thickness}~b) that is robust to large variations in the parameter space~\cite{supp}. We conclude that the charge in 2D is system-dependent. However, we stress that this charge is in good metallic contact with the superconductor, which makes it very difficult to probe it experimentally.  

A system that bypasses this issue is the proximitized 3D~TI, shown in Fig.~\ref{fig:3D_TI_with_vortex}. The normal state is described by the single particle Hamiltonian~\cite{fu_topological_2007, hasan_colloquium_2010},
\begin{equation}\label{eq:3DMTI_Hamiltonian}
H^{\textup{3D}}_0 = \sum_{i=x,y,z}t \sin k_i \tau_z\sigma_i + M({\mathbf{k}}) \tau_x + \beta({\bf r}) \sigma_z -\mu .
\end{equation}
Similar to the 2D case in Eq.~\eqref{eq:2D_2band_Hamiltonian}, this Hamiltonian also features spin-momentum coupling. Here, $\tau$ represents the orbital degrees of freedom,  $M({\mathbf{k}})=M_0+M_2\sum_{i=x,y,z}(1-\cos k_i)$, and $\mu$ is the chemical potential. This model describes a 3D~TI material, e.g. (Bi,Sb)$_2$Te$_3$ or Bi$_2$Se$_3$~\cite{zhang_topological_2009}.
We also allow for a position dependent magnetization, $\beta({\bf r})$.  It can be induced using the ferromagnetic proximity effect~\cite{katmis_high-temperature_2016,grutter_magnetic_2021} or we can use intrinsic magnetic TIs formed by magnetic doping (with e.g., V or Cr)~\cite{chang_experimental_2013, chang_high-precision_2015, ou_enhancing_2018} or formed as van der Waals structures such as MnBi$_2$Te$_4$~\cite{li_intrinsic_2019,rienks_large_2019,deng_quantum_2020, jansen_josephson_2024}.
We note that our results do not depend on the particular origin of the magnetization.

\begin{figure}
    \centering
    \includegraphics[width=\columnwidth]{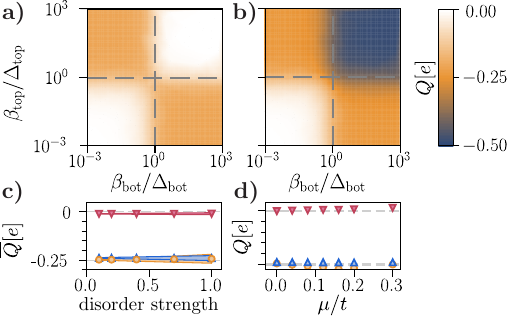} 
    \caption{\textbf{Robustness in a FMI/3D~TI/S heterostructure.}
     \textbf{a})  Center charge localized within a radius of $R+2a$ around the vortex core, and \textbf{b)} total charge as a function of the ratio of $\beta$ over $\Delta_0$ on the top and bottom surface of the 3D~TI.
    \textbf{c)} The ensemble ($N_{\rm ensemble}=1000$) averaged accumulated bottom (red down triangle), top (blue up triangle), and total (orange circle) charge as a function of chemical potential disorder strength normalized by the hopping strength in panel. Shaded colors correspond to one standard deviation from their mean values. \textbf{d)} Charge (same markers as in panel \textbf{c}) as a function of the chemical potential.} 
    \label{fig:3D charge robustness}
\end{figure}

When a vortex is introduced in the system and $\Delta_0 > |\beta|$ on a surface, it is well known that this surface hosts one MZM bound to the vortex core~\cite{fu_superconducting_2008}. 
Surprisingly, we find that the charge bound to this MZM vanishes. 
On the other hand, if $\Delta_0 < |\beta|$ on, say, the top surface, a $-e/4$ charge bound to the same vortex localizes where the flux exits, as illustrated in Fig.\ref{fig:increasing_thickness}~c. 
We plot the system-size dependence of this charge in Fig.~\ref{fig:increasing_thickness}~d. We find that the charge exponentially approaches its fractionally quantized value. Moreover, the induced charge is stable under variations of the magnetization as well as the superconducting gap, as shown in Fig.~\ref{fig:3D charge robustness}~(a-b). As a final check, we studied the effects of chemical potential (disorder) on this quantization in Fig.~\ref{fig:3D charge robustness}. We demonstrate that the quantization is extremely robust to the presence of disorder up to strengths comparable to the bandwidth. 

The separation of the MZM and fractional charge in 3D can be understood by considering the case where the 3D~TI is sandwiched between two magnetic insulators.  In the 2D limit, the height of the 3D~TI goes to zero and the top and bottom surfaces
hybridize. In this limit the two surfaces together form a QAH insulator with Chern number one~\cite{zhang_crossover_2010, asmar_topological_2018}. Laughlin's argument~\cite{laughlin_quantized_1981} then implies that a charge $-e/2$ is accumulated per flux $\Phi_0^{\text{sc}}$. Increasing the height of the TI does not change the total Chern number and the associated charge will be equally shared between the top and bottom surfaces due to symmetry. Hence a charge $-e/4$ is accumulated on each surface per $\Phi_0^{\text{sc}}$, demonstrating the "half Chern number" of the single surface of a 3D magnetic TI~\cite{nomura_surface-quantized_2011, grauer_scaling_2017}.

Increasing $\Delta_0$ beyond $\beta$ for the bottom surface, we fractionalize a fermion into two non-Abelian Ising anyons. The vortex-bound Ising anyon has a topological spin charge, following from Eq.~\eqref{eq: BP-N}, of $e/4$. However, this charge gets canceled by the opposite $-e/4$ charge originating from the half Chern number response to the vortex. One says that the Abelian anyon with charge $-e/4$ "fuses" with the non-Abelian Ising anyon. This forms $\bar \gamma$, a chargeless ($s=0$) non-Abelian anyon on the bottom surface and leaves a second isolated Abelian anyon of charge $-e/4$ at the top surface, as can be seen in Fig.~\ref{fig:increasing_thickness}~c. Finally, in the 2D limit (Fig~\ref{fig:increasing_thickness}~a), this top Abelian and bottom non-Abelian anyon also fuse to form $\gamma$, which has charge $-e/4$ ($s\ne0$).

We note that the topological spin is the underlying reason for the difference between our result that the charge at the S/3D~TI interface vanishes with that of Nogueira et al.~\cite{nogueira_fractional_2018}. In that work, the authors predicted an $-e/4$ Witten-effect charge at the S/3D~TI interface. However, since the effective field theory in Ref.~\cite{nogueira_fractional_2018}, did not explicitly include the MZM, and its topological spin,  the cancellation described above was not visible.

The fundamentally different behavior of MZMs in effective 2D systems and genuine 3D systems allows for a possible experimental control knob of their topological spin. The height of a 3D~TI can be controlled in a device geometry. Hence, it is conceivable to build devices where the braiding operation takes place effectively in 2D (top and bottom surfaces hybridize) or in 3D (top and bottom surfaces independent). If the MZMs are braided in effectively 2D we obtain the Abelian phase of $e^{-i\pi/4}$ seen in Fig.~\ref{fig:increasing_thickness}~(a-b). 
On the other hand, braiding in a 3D heterostructure, in the region with the proximity configuration as in Fig.~\ref{fig:3D_TI_with_vortex}, the Abelian  contribution is zero (Fig.~\ref{fig:increasing_thickness}~(c-d)). This is how one could manipulate the exchange statistics as well as the topological spin of the MZMs.

\begin{figure}
    \centering
    \includegraphics[width=\columnwidth]{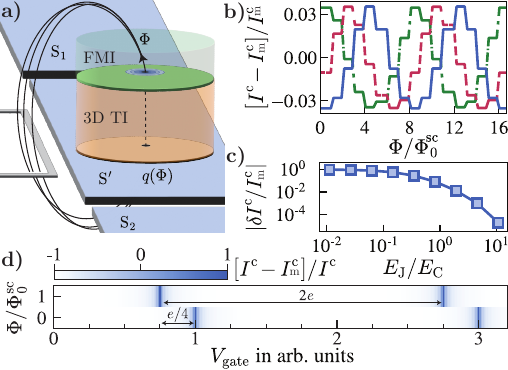} 
    \caption{\textbf{Experimental signatures.} \textbf{a)} A double Josephson junction set-up of two superconductors $\mathrm{S_{1,2}}$ coupled via a superconducting charge island $\mathrm{S}^\prime$' consisting of the S/3D~TI/FMI heterostructure in Fig.~\ref{fig:3D_TI_with_vortex}. The island gets charged through a flux modulation line and optionally an electrostatic gate.
    \textbf{b)} Mean subtracted critical current response as a function of the number of admitted vortices on the island for $q_{\mathrm{gate}}$ = 0 (solid blue line), 0.5e (dashed red line), and 1e (dash-dot green line), here $E_{\mathrm{J}}/E_{\mathrm{C}}=1.25$.
    \textbf{c)} Strength of oscillating critical current signal Eq.~\eqref{eq:total charge interference}, relative to its mean as a function of the $E_\mathrm{J}/E_\mathrm{C}$.
    \textbf{d)} Relative critical current signal as a function of the function of the number of admitted vortices and $V_\mathrm{gate}$ (which changes $q_\mathrm{gate}$), here $E_{\mathrm{J}}/E_{\mathrm{C}}=10^{-2}$.}
    \label{fig:single_arm_IC_quantized}
\end{figure}

We now focus on a concrete experimental proposal to measure the fractional charge of the vortex-bound anyons. 
We consider a double Josephson junction~\cite{reznik_question_1989,van_wees_aharonov-bohmtype_1990,friedman_aharonov-casher-effect_2002,hassler_anyonic_2010,elion_observation_1993,bell_spectroscopic_2016, de_graaf_charge_2018, randeria_dephasing_2024}, where Josephson vortices moving around a charged island interfere picking up phase differences depending on the charge on the island. Known as the Aharonov–Casher effect~\cite{aharonov_topological_1984}, this modulates the critical current as a function of charge. In our experimental proposal (Fig.~\ref{fig:single_arm_IC_quantized}~a), we place the 3D~TI on top of the center 
superconductor S$^\prime$, forming the  heterostructure in Fig.~\ref{fig:3D_TI_with_vortex}.
We couple S$^\prime$ to two superconducting leads $\text{S}_1$ and $\text{S}_2$ creating two Josephson junctions with phases $\phi_1,\phi_2$. A magnetic field is applied via a nearby flux modulation line, inducing vortices on S$^\prime$. As shown above, S$^\prime$ screens the vortex-induced charge in heterostructure, acquiring a charge $e/4$ per adimitted vortex. The top surface of the 3D~TI, which we assume to be far from S$^\prime$, carries the remaining $-e/4$. We call this (fractional) charge induced on S$^\prime$ by the vortices $q(\Phi)$, which is the property that we want to measure. 

To model the charge dynamics of S$^\prime$, we adopt a Cooper-pair box model in which the charge on the island is given by $2e \hat{n}$, where $\hat{n}$ is the operator that counts the number of Cooper pairs. We denote the superconducting phase on the island  $\theta$ and define the phase operator, $\exp \textstyle(i\hat{\theta}) |\theta'\rangle = \exp (i \theta ')|\theta'\rangle$. Here $\vert\theta'\rangle$ is the BCS ground state of the island with phase $\theta'$. It is well known that  $\exp \textstyle(i\hat{\theta})$ increases the number of Cooper pairs by one, hence 
$[\hat n, \exp \textstyle (i\hat{\theta})] = \exp \textstyle(i\hat{\theta})$. 
In order to describe the coupling to the fermions in the topological material we replace the phase $\phi_0$ with the operator $\hat{\theta}$ in Eq.~\eqref{Eq:BCS},
\begin{equation}
   \mathcal{H}_{\rm F}= \sum_\mathbf{k}\psi^\dagger_{\mathbf{k}} 
   H_0({\mathbf{k}})\psi^{\phantom{\dagger}}_{\mathbf{k}} + \sum_{j} 
   \Delta_0 e^{i\varphi_j +i \hat{\theta}}\psi_{j\uparrow}\psi_{j\downarrow} + \text{h.c.}. 
\end{equation}
The term $\exp\small(i\hat{\theta}\small)\psi_{j\uparrow}\psi_{j\downarrow}$ now describes the creation of a Cooper pair on S$^\prime$, while removing a pair of fermions from the topological material at site $j$. Notice that, in this model, the total charge is conserved, making it ideal for describing fractional charges in a superconducting setting. 

We describe the island's charging energy using a capacitive model,
\begin{equation}
    \mathcal{H}_{\rm C}=(\hat{Q}-q_{\rm gate})^2/2C,
\end{equation}
where the charge operator $\hat{Q}$ is the sum of all the charge on S$^\prime$ as well as all the fermionic charge in good metallic contact with S$^\prime$,  
\begin{equation}\label{eq:total charge operator}
  \hat{Q} = 2e \hat{n}+e\sum_i f_i  \psi^{\dagger}_i \psi^{\phantom{\dagger}}_i. 
\end{equation}
Here $f=1$ for sites in good metallic contact with S$^\prime$, and $f=0$ otherwise. Hence, a 2D material in contact with the island will have $f=1$ everywhere. In contrast, for the 3D setup in Fig.\ref{fig:single_arm_IC_quantized}~a, the bottom and side surface states will have $f=1$, while the top surface, which we assume to be sufficiently separated from S$^\prime$ by an insulating layer~\footnote{We note that if the insulating layer is not thick enough, this model needs to be modified. Choosing a boundary which is fuzzy, i.e.~$f$ changes smoothly from $1$ to $0$ within the insulating layer, does not change our results.}, will have $f=0$. 
The charge $q_{\rm gate}$ is induced due to the electrostatic environment, which could optionally be controlled using an electrostatic gate $V_\mathrm{gate}$. It simply shifts the expectation value of $\hat{Q}$. 

The Hamiltonian $\mathcal{H}_0=\mathcal{H}_{\rm F}+ \mathcal{H}_{\rm C}$ commutes with the total number operator $\hat{N}_{\rm tot}=2\hat{n} + \hat{N}_{\rm B}+\hat{N}_{\rm T}$, where $\hat{N}_{\rm B}=\sum_i f_i  \psi^{\dagger}_i \psi^{\phantom{\dagger}}_i$ and $\hat{N}_{\rm T}=\sum_i  (1-f_i) \psi^{\dagger}_i \psi^{\phantom{\dagger}}_i$. Thus,
in two-dimensional systems, where $\hat{N}_{\rm T} = 0$, we work in simultaneous eigenstates of $\mathcal{H}_0$ and $\hat{N}_{\rm tot} \equiv \hat{Q}$. These number eigenstates are given by
\begin{equation}
    \vert N' \rangle= \int d\theta {\rm e}^{iN'\theta} \vert \theta\rangle\otimes\vert {\rm BCS}_\theta\rangle.
\end{equation}
For the 3D systems we consider here, although $\{\vert N' \rangle\}$ are not simultaneous eigenstates, we assume that the low energy dynamics is within the subspace spanned by $\{\vert N' \rangle\}$. This assumption requires 
the (magnetic insulating) top layer to have a large energy gap so that quasiparticle excitations involving the top layer are suppressed.
In 2D, the fermionic charge and induced image charge in S$^\prime$ are in good metallic contact and thus cancel, but in 3D the fermionic charge in the top layer is isolated and induces, additional to $q_\mathrm{gate}$ an image charge on S$^\prime$,
\begin{equation}
    q_{\rm gate}\rightarrow q(\Phi) = q_{\rm gate}+ e\langle N'\vert\hat{N}_T\vert N'\rangle 
\end{equation}
This is set by the flux $\Phi$ (number of vortices) through the heterostructure. 
This highlights the crucial difference between a 2D and 3D set-up.
As shown above, each admitted vortex induces $e/4$ in S$^\prime$, giving $q(\Phi) =q(0) +\frac{e}{4} \Phi/\Phi^{\rm SC}_0$. 

We now couple the Cooper pair box to the two superconducting leads. The Hamiltonian becomes
\begin{equation}\label{eq:ham single arm}
H= \frac{e^2}{2C} \left (-i \frac{\partial}{\partial \theta}-\frac{q(\Phi)}{e}\right )^2
- E_{\mathrm{J1}}\cos\phi_1 -E_{\mathrm{J2}}\cos\phi_2,
\end{equation}
where $\theta=(\phi_1-\phi_2)/2$, and we define $\varphi_{\mathrm{T}}= \phi_1+\phi_2$ as the classical (controllable) target flux. For ease of discussion~\cite{supp_interference}, we consider the junctions to have equal Josephson energies $E_{\mathrm{J}}$.
We compute the critical current as the maximally carried supercurrent $I_{\text{sc}}(\varphi_\mathrm{T}) = \frac{dE_0}{d\varphi_\mathrm{T}}$, where $E_0(\varphi_\mathrm{T})$ is the ground state energy of $H$ given in Eq.~\eqref{eq:ham single arm}. The critical current response, shown in Fig.~\ref{fig:single_arm_IC_quantized}~b, has quantized values depending on the number of admitted vortices $n_\Phi= \Phi/\Phi^{\text{sc}}_{0}\in\mathbb{N}$. It is periodic with eight admitted vortices, directly relating the measured periodicity to the fractional charge response on the island. In the strong Josephson coupling limit $E_J > E_C=(2e)^2/2C$, we have
\begin{equation}\label{eq:total charge interference}
I^\mathrm{c} = I^\mathrm{c}_\mathrm{m} + \delta I^\mathrm{c} \cos\frac{\pi q(\Phi)}{e}.
\end{equation}
Hence, we see an oscillating signal of amplitude $\delta I^\mathrm{c}$ around a mean $I^\mathrm{c}_\mathrm{m}$. In Fig.~\ref{fig:single_arm_IC_quantized}~c, we study the relative strength of this signal and find that it monotonically increases towards the regime $E_\mathrm{J}<E_\mathrm{C}$, where charging energy dominates. Here, the periodicity of the signal remains the same, but its shape is distorted from a pure cosine function~\cite{bozkurt_josephson_2023}. 

The critical currents depend on the gate voltage; however, in the strong Josephson coupled regime, the values of the eight quantized critical current values are related in a gate invariant manner. We find that they satisfy a set of four universal equations that directly relate to the observation of the quantized fractional charge (thus the topological spin), 
\begin{equation}
\label{eq:universal critical current relations}
\begin{split}
        I^{\rm c}_{\mathrm{i}}  = (I^{\rm c}_{\mathrm{i+1}} - I^{\rm c}_{\mathrm{i+3}})/\sqrt{2},\qquad  \text{for i = 0,4,}\\
        I^{\rm c}_{\mathrm{i}}  = (I^{\rm c}_{\mathrm{i-1}} + I^{\rm c}_{\mathrm{i+1}})/\sqrt{2},\qquad  \text{for i = 2,6,}
\end{split}
\end{equation}
with $I^{\mathrm{c}}_i$ being the $i$th plateau appearing in sequential order starting from $I^ {\rm{c}}_0$ being the lowest plateau. 

In the regime $E_\mathrm{J}<E_\mathrm{C}$, where the charging energy dominates, the universal supercurrent relations break down. However, the optional control of a gate voltage offers an additional method to probe the fractional charge. When we sweep the gate voltage and thus $q_\mathrm{gate}$ in Eq.~\ref{eq:total charge interference}, and simultaneously increase the induced flux $\Phi$ and thus $q(\Phi)$, we obtain the map shown in Fig.~\ref{fig:single_arm_IC_quantized}~d. From the shift in the gate sweeps, we can directly read off the fractionally quantized charge. We stress that the effect is observable for all $ E_{\mathrm{J}}/ E_{\mathrm{C}}$ regimes, but the peak of the critical current becomes narrower in the charging energy dominant regime, making read-out more straightforward. The horizontal axis is a voltage and not a charge, but if one normalizes the gate sweep by one period (corresponding to $2e$ or depending on the experimental details $e$) the fractional charge signature is recovered.
This approach is similar to recent experiments probing the fractional charge in fractional quantum Hall states~\cite{de_c_chamon_two_1997,nakamura_direct_2020,feldman_fractional_2021,kundu_anyonic_2023}. Although in our setup, the signal is in the critical supercurrent response rather than in the dissipative current response.

In conclusion, we have established a link between the topological spin and fractional charge in vortex-bound Majoranas, which allowed us to probe their topological spin across different systems. This revealed that the topological spin in a S/3D~TI/FMI heterostructure can be manipulated in an MZM exchange operation by leveraging device geometries. The control over the topological spin of Majoranas, together with control of Josephson vortices  could open the door to new topological braid operations.

Finally, we proposed an experimental setup that is capable of detecting the topological spin of Majoranas by detecting the presence of a fractionally quantized charge of $-e/4$ in both the weak and the strong Josephson coupling regime. The proposed flux-gated vortex interference experiment can be straightforwardly implemented with currently available technology~\cite{schuffelgen_selective_2019,brevoord_phase_2021,park_vortex-parity-controlled_2026},
bringing the detection of this elusive topological property within reach.

{\textit{Acknowledgments}}. \.Inan\c{c} Adagideli and Stijn de Wit thank Carlo Beenakker for inspiration and fruitful discussions.

\let\oldaddcontentsline\addcontentsline
\renewcommand{\addcontentsline}[3]{}
\bibliographystyle{apsrev4-2}
\bibliography{arxiv_references}

\let\addcontentsline\oldaddcontentsline

\clearpage
\newpage
\pagenumbering{arabic}  
\setcounter{equation}{0}
\setcounter{figure}{0}
\setcounter{section}{0}

\renewcommand{\thesection}{S\arabic{section}}
\renewcommand{\theequation}{S\arabic{equation}}
\renewcommand{\thefigure}{S\arabic{figure}}

\clearpage
\onecolumngrid
\begin{center}
\textbf{\large Supplemental Material: } \\ 

\vspace{.1cm}
\textbf{\large Manipulating the topological spin of Majoranas}\\
\vspace{.5cm}
Stijn R. de Wit$^{1,*}$, Emre Duman$^{1,2,*}$, A. Mert Bozkurt$^{3,4}$, Alexander Brinkman$^{1}$, \.Inan\c{c} Adagideli$^{1,2,5,\dagger}$\\
\vspace{.1cm}
\small{ 
$^{1}$\it{MESA+ Institute for Nanotechnology, University of Twente, The Netherlands} \\
$^{2}$\it{Faculty of Engineering and Natural Sciences, Sabanci University, Istanbul, Turkey} \\
$^{3}$\it{QuTech, Delft University of Technology, Delft 2600 GA, The Netherlands} \\
$^{4}$\it{Kavli Institute of Nanoscience, Delft University of Technology, P.O. Box 4056, 2600 GA Delft,
The Netherlands} \\
$^{5}$\it{T\"UB\.ITAK Research Institute for Fundamental Sciences, Turkey}\\
$^*$\it{These authors contributed equally}}\\
$^*$\it{Correspondence to: adagideli@sabanciuniv.edu}
\end{center}
\twocolumngrid
\tableofcontents

\section{Berry Connection from Read's Formula}
\label{app:berry_connection_details}
In this section, we walk through the details of the Berry connection, and demonstrate its equivalence to the expectation of the number operator numerically. 
We want to compute the Berry connection associated with the operation (in Eq.~\eqref{eq:berry connection phi0} of the manuscript) of braiding a pair of Majoranas around each other by cycling $\phi_0 \to \phi_0+2\pi$ of the BCS ground state in Eq.~\eqref{eq:BCS} of the manuscript.

The evaluation of the Berry connection is more straightforward in the Thouless representation of the BCS wavefunction expressed in fermionic quasiparticle annihilation operators $\{\hat\psi_i\}$~\cite{read_non-abelian_2009},
\begin{align}\label{eq:HFB}
|\Omega(\phi_0)\rangle=A \, \exp \left(\sum_{i<j} Z_{i j} \psi_i^{\dagger} \psi_j^{\dagger}\right)|0\rangle,
\end{align}
with $\ket{0}$ being the vacuum state. Here the skew-symmetric matrix $Z = i\sigma_y V^* U^{-1}$ and normalization $A=\sqrt{|\det U|}$ are related to the positive energy eigenvectors $U_{ij}=u_j(i)$
and $V_{ij}=v_j(i)$ of the discretized mean field Bogoluibov de-Gennes (BdG) Hamiltonian,
\begin{align}\label{eq:BdG Hamiltonian app}
\left(\begin{array}{cc}
{H_0}_{ij} & \Delta_0 e^{i\varphi_i}\delta_{ij} \\
 \Delta_0 e^{-i\varphi_i}\delta_{ij}  & -\sigma_y  {H_0}_{ij}^* \sigma_y
\end{array}\right)
\left(\begin{array}{cc}
u_n(j)\\
v_n(j)
\end{array}\right) 
=E_n\left(\begin{array}{cc}
u_n(i)\\
v_n(i)
\end{array}\right) 
.
\end{align}
Having calculated the complete set of eigenvectors of the BdG Hamiltonian in Eq.~\eqref{eq:BdG Hamiltonian app}, we now evaluate the 
Berry connection in Eq.~\eqref{eq: BP-N} using  Read's formula~\cite{read_non-abelian_2009}
\begin{equation}\label{eq:reads-formula}
\mathcal{A} = \frac{i}{4}\operatorname{Tr}\left( \left[Z^\dagger \frac{\partial Z}{\partial \phi_0}-\frac{\partial Z^\dagger}{\partial \phi_0}Z\right]\left(1+Z^\dagger Z\right)^{-1} \right).
\end{equation}
We compare this to the expectation value of the number operator  $\hat{N}=\sum_{i,\sigma}\psi^\dagger_{i,\sigma}\psi^{\phantom{\dagger}}_{i,\sigma}$.

\begin{figure}
    \centering
    \includegraphics[width=\columnwidth]{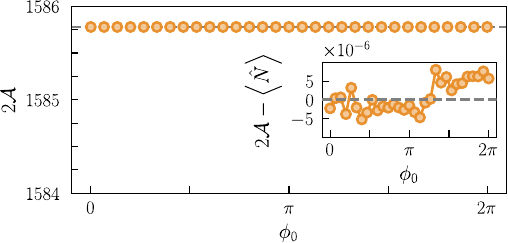}   
    \caption{\textbf{Fractional charges and Berry connection.} Twice the Berry connection $2\mathcal{A}$ (orange circles) and the number operator expectation value $\langle \hat N\rangle$ (dashed gray line) as a function of superconducting phase $\phi_0$, for the four band 2D model.
    Inset: deviation of ${2\mathcal{A}}$ from $\langle \hat N\rangle$.} 
    \label{fig:berry connection}
\end{figure}

We first demonstrate the relation, $2\mathcal{A}=\langle \hat N\rangle$, from the manuscript, for the 2D four band model by evaluating the Berry connection using Read's formula and comparing it to the total Fermi-sea charge. 
As we show in Fig.~\ref{fig:berry connection}, twice the Berry connection indeed approximately equal to the fractional part (modulus 2) of the number expectation value. The relative error between the two is below $10^{-8}$. The error can be attributed to the accumulated numerical error due to matrix inversion and finite difference while evaluating the partial derivatives in Eq.~\eqref{eq:reads-formula}.

\section{Details of the tight-binding calculations}
Here we provide details of the real space tight-binding implementation of the lattice Hamiltonians from the manuscript. In addition, for the 2D models, we compare the two band QAH to the four band model, which we interpret as the 2D limit of the 3D~TI. The lattice Hamiltonians are of the form Eq.~\eqref{eq:BdG Hamiltonian app}, but their exact implementation depends on the single-particle Hamiltonian $H_0$. The real space formulation of $H_{ij}^{\rm BdG}$ is obtained by the mappings: 
$\sum_{i=x,y}\sin{k_i} \to 1/2i\sum_{\mathbf{r},\mathbf{ \delta}} \ket{\mathbf{r} + \boldsymbol{\delta} }\bra{\mathbf{r}} + \text{h.c.}$, and 
$\sum_{i=x,y}\cos{k_i} \to 1/2\sum_{\mathbf{r},\mathbf{ \delta}} \ket{\mathbf{r} + \boldsymbol{\delta} }\bra{\mathbf{r}} + \text{h.c.}$.  
We give the details of the 2D and 3D models separately. In table~\ref{table:TB-params} we summarize the specific parameters used in each figure in the manuscript and in this supplemental material.

\subsection{Two-dimensional models}
The 2D geometry is defined in the discrete coordinates, that is, $\left\{(x,y): x, y \in \mathbb{N}\,\text{ and }\, x^2 + y^2\leq r^2   \right\}$. Accordingly, the discrete displacement vector $\boldsymbol{\delta}$ is the distance between neighboring lattice sites and in 2D takes values from the set  $\{a\hat{\boldsymbol{x}},\, a \hat{\boldsymbol{y}}\}$. Here, $a$ is the lattice spacing, which we set to 1 throughout. We first consider the two band 2D lattice Hamiltonian \eqref{eq:2D_2band_Hamiltonian} from the manuscript, it reads,  
\begin{widetext}
\begin{align}
   H^{\text{2D}} = & \sum_{\boldsymbol{r}}\left(\begin{array}{cc}
        \left(\beta_0+2\beta_2\right)\sigma_z + \mu & \Delta_0 e^{i\phi(\mathbf{r})} \\
        \Delta_0 e^{-i\phi(\mathbf{r})} & \left(\beta_0+2\beta_2\right)\sigma_z - \mu
\end{array}\right)\ket{\mathbf r} \bra{\mathbf r} \nonumber\\
&  \qquad \qquad  +\sum_{ \boldsymbol{\delta},\mathbf{r}}\left(\begin{array}{cc}
        \exp{i\int_{\mathbf{r}}^{\mathbf{r}+\boldsymbol{\delta}} \boldsymbol{A}  \cdot \boldsymbol{dl}}  & 0 \\
        0 & \exp{-i\int_{\mathbf{r}}^{\mathbf{r}+\boldsymbol{\delta}}\boldsymbol{A}\cdot \boldsymbol{dl}}
    \end{array}\right)\left(\frac{it}{2} \boldsymbol{\delta} \cdot \boldsymbol{\sigma}+ \frac{\beta_2}{2}\sigma_z\right)\ket{\mathbf{r}+\boldsymbol{\delta} }\bra{\mathbf{r}}+\mathrm{h.c.}.
\end{align}
\end{widetext}
The scalar phase function $\phi({\boldsymbol r})$ is the azimuthal angle around the vortex core penetrating through the center-most plaquette. The vector potential describes a magnetic field $\boldsymbol{B}=\Theta(R-{\boldsymbol{\rho}})\hat{{\boldsymbol z}}$  where $\boldsymbol{\rho}= x\hat{\boldsymbol{x}}+y\hat{\boldsymbol{y}}$ is the radial distance from the vortex core, and $R$ marks the radius of the circular region where the magnetic field acts. Neither depends on the $z$ coordinate, so the vector potential reads, 
\begin{equation}
\boldsymbol{A} =
\begin{aligned}\begin{cases}
\Phi\frac{\rho}{2\pi R^2}\hat{\boldsymbol{\phi}}, \quad \rho<R\\
\Phi\frac{1}{2\pi \rho}\hat{\boldsymbol{\phi}},  \quad \rho>R.\\
\end{cases}\end{aligned}
\end{equation}
Where $\Phi = n\Phi_0$ is the total flux being an integer multiple of the superconducting flux quanta, and  $\hat{\boldsymbol{\phi}} = -y \boldsymbol{x}+ x \boldsymbol{y}$ is the azimuthal angle around the vortex core at ${\bf r}_0 = (0,\,0)$. 

We now study the two band model's dependence on the parameters, $\beta_0,\Delta_0$, and $\beta_2$, introduced in the manuscript. The momentum dependent regularization term $\beta_2\sum_{i=x,y}(1-\cos k_i)$ is required in the model to gap out the secondary Dirac cones at the edge of the Brillouin zone. In absence of $\Delta_0$ and $\mu$, the system is in the QAH phase\cite{qi_chiral_2010} whenever $\beta_0/\beta_2<0$. Without loss of generality, we choose $\beta_2<0$. For $|\beta_0|>\Delta_0>0$, the magnetization dominates. For $\beta_0>0$, the system is in a $|C|=2$ topological superconductor phase. while for $\beta_0<0$, the system is in a topologically trivial state. It transitions to a Chern number $|C|=1$ topological superconductor when $\Delta_0>|\beta_0|$. 

\begin{figure}
    \centering
    \includegraphics[width=\columnwidth]{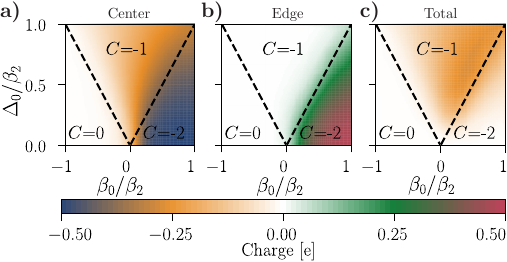}
    \caption{\textbf{Two band 2D model charges as a function of $\beta$ over $\Delta_0$}.
    The real-space localization of the charge in the center \textbf{(a)} within a radius of $R+2a$ around the vortex core, at the edge \textbf{(b)} within $2a$ from the edge, and the total charge \textbf{(c)}. The black dotted lines indicate the topological phase transition lines~\cite{qi_chiral_2010} between regions of different Chern numbers. The color bar is shared across the figure.} 
    \label{fig:phase-space-2d_2band}
\end{figure}

In Fig.~\ref{fig:phase-space-2d_2band}, we show the induced Fermi-sea charge as a function of $\beta_0$ and $\Delta_0$ relative to $\beta_2$ and a fixed $\mu=0$. We see that the induced charge globally follows the phase transition lines of the bulk system mentioned above(dotted lines). Hence there is a charge of $0$ that localizes around the vortex' core (panel a) where $C$ vanishes, and $-e/2$ where $C=-2$. In the $|C|=2$ topological superconductor phase, an edge charge is also  present (panels b), which counters the center charge (panel c). In contrast,  for the $|C|=1$ topological superconductor phase, the edge modes are charge neutral and the total charge is given by the center charge. Note that we don't find this charge to be quantized over a large stable phase space. Especially near the phase transitions, where the gaps diverge, the localized charge is not stable. We ascribe this to finite size effects. 

A more realistic 2D model, where the regularization does not break time-reversal symmetry, is a four band model that describes two (tunnel) coupled TI surfaces,
\begin{equation}\label{eq:2D_Hamiltonian_4band}
    \tilde H^{\text{2D}} = \sum_{i=x,y}t \sin k_i \tau_z \sigma_i + m(\boldsymbol{k}) \tau_x + \sum_{j=\mathrm{top},\mathrm{bot}}\beta_j\tau_j\sigma_z -\mu.
\end{equation}
Here, the first, second, and third term represents spin-momentum coupling,
the regularization term $m(\boldsymbol{k})=m_0+m_2\sum_{i=x,y}(1-\cos k_i)$ which mixes the two surface states, top surface and bottom surface, indexed with the $\tau$ Pauli matrices via $\tau_{\mathrm{top}}=(\tau_0+\tau_z)/2$ and $\tau_{\mathrm{bot}}=(\tau_0-\tau_z)/2$, and magnetization respectively. This model describes the thin-limit of a 3D~TI where two opposite surface are tunnel-coupled via $m(\boldsymbol{k})$. This regularization does not break time-reversal symmetry. 

\begin{figure}
    \centering
    \includegraphics[width=\columnwidth]{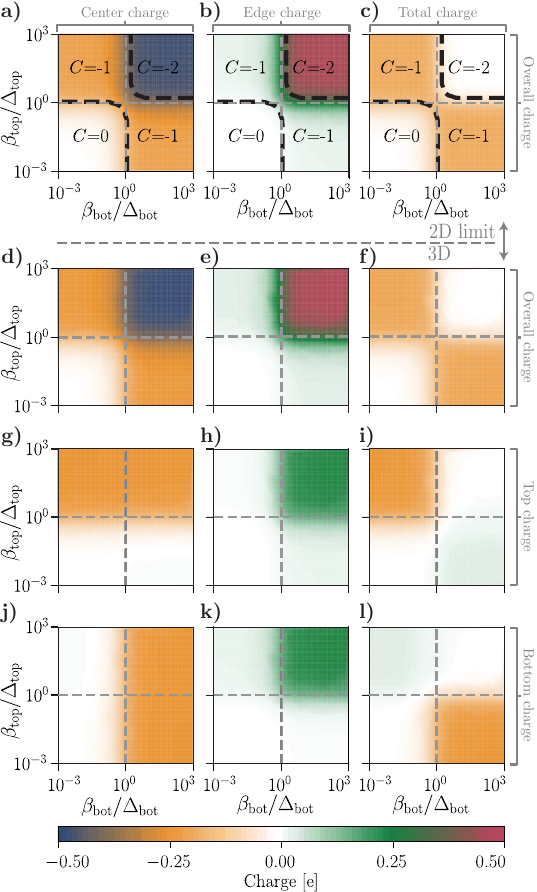}
    \caption{\textbf{Charge phase space for the four band model in 2D and 3D}.
     \textbf{a-c)} 2D four band model: charges as a function of the ratio of $\beta$ over $\Delta_0$ on the top and bottom surfaces. Sub-figures are grouped in columns: the real-space localization of the charges within a radius of $R+2a$ around the vortex core in the center (left column), within $2a$ from the edge (middle column), and total charge (right column). The black dotted lines indicated domains of constant Chern number, calculated using Fukui-Hatsugai-Suzuki method\cite{fukui_chern_2005}.
     \textbf{d-l).} 3D four band model: in rows we plot the charge for the overall (\textbf{d-f}) system, the top two layers in $z$ (\textbf{g-i}), and bottom two layers in $z$  (\textbf{j-l}). The same labels and column structure (center, edge, total) holds as for the above row. The color bar is shared across the figure. 
    } 
    \label{fig:phase-space-2d_and_3d}
\end{figure}

If we place this system in proximity contact to an s-wave superconductor, we can tune the relative strength of $\beta/\Delta_0$ on the $\tau_\mathrm{top}$ and $\tau_{\mathrm{bot}}$ blocks. When $\Delta_0$ dominates on both $\tau$ blocks, a $\Phi_0$ vortex simply binds a pair of two MZMs localized on either $\tau_{\mathrm{top},\mathrm{bot}}$ block. However, in the limit where the proximity gap $\Delta_0$ only dominates in one block, we get a situation where a single MZM and $-e/4$ charge localize at the vortex core. One can interpret this situation as the 2D limit of the 3D~TI. In Fig.~\ref{fig:phase-space-2d_and_3d}, we study the charge response dependency on the $\beta$ and $\Delta_0$ gaps. We vary their relative gap size on each (top and bottom) surface. The picture is very simple; whenever $\beta$ dominates over $\Delta_0$ locally on a surface, a charge of $-e/4$, corresponding to a Chern number $C=-1$, localizes around the vortex' center (panel a). In the absence of $\Delta_0$, an edge charge is present (panel b) that screens the center charge, resulting in a total charge of $-e/4$ (panel c). This quantized total charge is stable over orders of magnitude.

\subsection{Three-dimensional TI model}
The geometry for 3D is the natural extension of the 2D case and it's defined in the discrete coordinates $\left\{(x,y,z): x, y, z \in \mathbb{N}\, \text{ and }\, x^2 + y^2\leq r^2\, \text{ and }\, 0\leq z < h  \right\}$. Accordingly, the discrete displacement vector $\boldsymbol{\delta}$ is the distance between neighboring lattice sites in 3D and is in the set $\{a\hat{\bf{x}},\, a \hat{\bf{y}},\, a\hat{\bf{z}}\}$. The 3D~TI lattice Hamiltonian \eqref{eq:3DMTI_Hamiltonian} from the manuscript, similar to the 2D case, reads,
\begin{widetext}
\begin{align}
    H^{\text{3DTI}} = & \sum_{\boldsymbol{r}}\left(\begin{array}{cc}
        \beta(\boldsymbol{r})\sigma_z + (M_0+2M_2)\tau_x + \mu & \Delta_0(\boldsymbol{r}) e^{i\phi(\boldsymbol{r})} \\
        \Delta_0(\boldsymbol{r}) e^{-i\phi(\boldsymbol{r})} & \beta(\boldsymbol{r})\sigma_z +  (M_0+2M_2)\tau_x - \mu
    \end{array}\right)\ket{\boldsymbol{r}}\bra{\boldsymbol{r}} \nonumber  \\
    &  \qquad \qquad  +\sum_{ \boldsymbol{\delta},\boldsymbol{r}}\left(\begin{array}{cc} \exp{i\int_{\boldsymbol{r}}^{\boldsymbol{r}+\boldsymbol{\delta}} \boldsymbol{A}  \cdot \boldsymbol{dl}}  & 0 \\
        0 & \exp{-i\int_{\boldsymbol{r}}^{\boldsymbol{r}+\boldsymbol{\delta}}\boldsymbol{A}\cdot \boldsymbol{dl}}  \end{array}\right)\left(\frac{it}{2} \boldsymbol{\delta} \cdot \boldsymbol{\sigma}+ \frac{M_2}{2}\tau_x\right)\ket{\boldsymbol{r}+\boldsymbol{\delta} }\bra{\boldsymbol{r}} + \mathrm{h.c.}.
\end{align}
\end{widetext}
Unlike the 2D case, in the 3D Hamiltonian the superconducting pair potential strength $\Delta_0 = \Delta_0(z)$ and the magnetization strength $\beta(\boldsymbol{r}) = \beta(z)$ are $z$ coordinate-dependent. 

We work with different $z$ dependency for $\beta$ and $\Delta_0$. When one dominates throughout the system, the exact form of their $z$ dependency does not matter ostensibly. The relevant case is when they dominate the other on opposing surfaces. We choose an exponentially decaying (from the surface where they dominate), so the form of these functions look like $f(z) = c e^{-z/\xi}$ if they dominate in the bottom surface, and $f(z) = c e^{-(h-z)/\xi}$ if they dominate at the top surface, where $\xi$ is a decay rate and $h$ is the height of the system. 

The 3D Fermi-sea charges as a function of the relative gap size of $\beta(z)$ and $\Delta_0(z)$ at the bottom ($z=0$) and top $(z=h)$ are shown in Fig.~\ref{fig:phase-space-2d_and_3d}~(d-l). Note that panels d and f also feature in Fig.~\ref{fig:3D charge robustness} in the manuscript. A charge of $-e/4$ localizes around the vortex core where $\beta$ dominates over $\Delta_0$. The crucial difference w.r.t. the four band 2D model in Eq.~\ref{eq:2D_Hamiltonian_4band} is that the localized charge is completely separated in real space from the MZM localized on the opposite surface where $\Delta_0$ dominates.

\begin{table}[h!]
\begin{tabular}{lllll}
\multicolumn{1}{l|}{Fig.}  & Sizes (r,R,h)   & Regular. & $\beta_\mathrm{bot},\beta_\mathrm{top}$ & $\Delta_\mathrm{bot},\Delta_\mathrm{top}$\\ \hline
\multicolumn{1}{l|}{2.a}   & (28,4,-)                 & $-1.35$           & $0.12$                                  & $0.61$         \\
\multicolumn{1}{l|}{2.b QAH} & ($r$,4, -)             & $-1.35$           & $0.12$                                  & $0.61$  \\
\multicolumn{1}{l|}{2.b 4 band} & ($r$,4, -)          & $(0,1.23)$        & $(1.00,0.01)$                           & $(0.60,0.01)$   \\
\multicolumn{1}{l|}{2.c}   & (18,4,6)                 & $(-1.5,1.65)$     & $(0,0.5)$                               & $(0.7,0)$          \\
\multicolumn{1}{l|}{2.d}   & ($r$,4,4)                & $(-1.5,1.65)$     & $(0,0.5)$                               & $(0.7,0)$          \\
\multicolumn{1}{l|}{3.a-b} & (12,3,-)                 & $(0,1.23)$        & 0 to $0.5m_2$                           &  0 to $0.5m_2$           \\
\multicolumn{1}{l|}{3.c}   & (8,3,4)                  & $(-1.5,1.65)$     & $(0,0.5)$                               & $(0.7,0)$          \\
\multicolumn{1}{l|}{3.d}   &(16,4,4)                 & $(-1.5,1.65)$     & $(0,0.5)$                               & $(0.85,0)$  \\
\multicolumn{1}{l|}{S1}    & (16,4,-)                 & $(0,1.23)$        & $(0.01,1.04)$                           & $(0.60,0.01)$          \\
\multicolumn{1}{l|}{S2}    & (15,3,-)                 & $-1.35$           & $-1.35$ to $1.35$                       & $0$ to $1.35$           \\
\multicolumn{1}{l|}{S3.a-c}& (12,3,-)                 & $(0,1.23)$        & 0 to $0.5m_2$                           &  0 to $0.5m_2$     \\
\multicolumn{1}{l|}{S3.d-l}& (10,3,4)                 & $(-1.5,1.65)$     & 0 to $0.5M_2$                           &  0 to $0.5M_2$     \\
\end{tabular}\caption{\textbf{Tight-binding parameters used in each figure.}. The regularization takes the form $\beta_2$ for the two band QAH model,  ($m_0,m_2$) for the 2D four band model, and ($M_0,M_2$) for the 3D four band model. For the 2D QAH model, there is no distinction between top and bottom for $\beta$ and $\Delta_0$. All sizes are in units of the lattice constant $a$, and all parameters in units of the in plane hopping parameter $t=1$. In 3D, $t_z=t/2$.}\label{table:TB-params}
\end{table}

\section{Critical current relationships}\label{app:vortex_interference}
\begin{figure}
    \centering
    \includegraphics[width=\columnwidth]{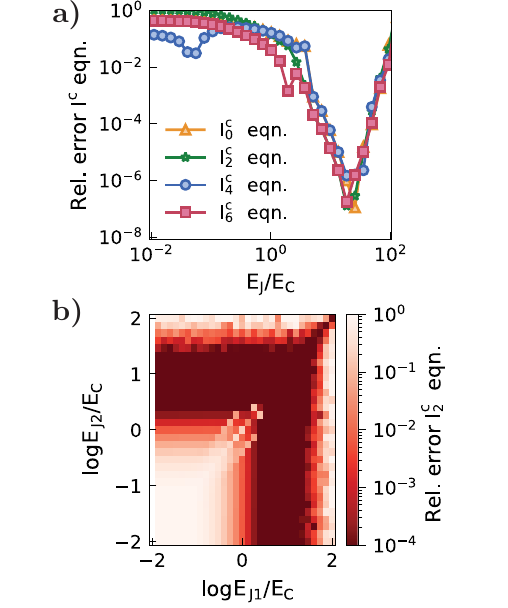} 
    \caption{\textbf{Relative error critical current relations.}
    \textbf{a)} Error relative to the amplitude of the critical current oscillation of the equations~\eqref{supp eq:universal critical current relations}, relating the universal quantization of critical current plateaus as a function of the Josephson energy relative to the capacitive energy. Here, $q_\mathrm{i}\approx 0.16$~e.
    \textbf{b)} 
    Heat map of the error relative to the amplitude of the critical current oscillation of the  $I^{\mathrm{c}}_2$ Eq.~\eqref{supp eq:universal critical current relations} relating the universal quantization of critical current plateaus as a function of the Josephson energies $E_{\mathrm{J1}}$ and $E_{\mathrm{J2}}$ normalized by $E_\mathrm{C} = (2e)^2/2C$. Here, $q_\mathrm{i}=0$.
    } 
    \label{fig:single_arm_IC_quantized_sup}
\end{figure}

In this section we provide the scaling of the relative evaluation error of the set of equations relating the quantized $I^\mathrm{c}$ plateaus from the main text. The universal relationship between the plateau values in the manuscript Eq.~\eqref{eq:universal critical current relations} are,
\begin{equation}\label{supp eq:universal critical current relations}
    \begin{cases}
        I^{\mathrm{c}}_0  = (I^{\mathrm{c}}_1 - I^{\mathrm{c}}_3)/\sqrt{2},\\
        I^{\mathrm{c}}_2  = (I^{\mathrm{c}}_1 + I^{\mathrm{c}}_3)/\sqrt{2},\\
        I^{\mathrm{c}}_4  = (I^{\mathrm{c}}_5 - I^{\mathrm{c}}_7)/\sqrt{2},\\
        I^{\mathrm{c}}_6  = (I^{\mathrm{c}}_5 + I^{\mathrm{c}}_7)/\sqrt{2}.
    \end{cases}
\end{equation}
Their evaluation error relative to the amplitude of the critical current oscillation is plotted in Fig.~\ref{fig:single_arm_IC_quantized_sup}~a for an increasing Josephson energy relative to the capacitive energy $E_\mathrm{C} = (2e)^2/2C$. The set of equations becomes fully resolvable in the strong Josephson coupling regime (relative error below $10^{-2}$) when $E_{\mathrm{J}}/E_C\gtrapprox 1$ and decreases further to $\sim10^{-8}$. The decrease in the relative error is haltered by a simultaneous decrease of the signal strength $\delta I^{\mathrm{c}}$, increasing the relative error back to unity. 

In the manuscript, we work, for simplicity, with assumption of equal Josephson energies, but this can be relaxed. We illustrate that in Fig.~\ref{fig:single_arm_IC_quantized_sup}~b by computing the evaluation error of $I^{\mathrm{c}}_2$ in Eq.~\eqref{supp eq:universal critical current relations} for $q_{\mathrm{i}}=0$ as a heat map of $E_\mathrm{J1}/E_{\mathrm{C}}$ and $E_\mathrm{J2}/E_{\mathrm{C}}$. The same behavior as for the equal Josephson energy case in Fig.~\ref{fig:single_arm_IC_quantized_sup}~a is observed. Now, however, the error is governed by the maximal value of $E_\mathrm{J1}$ and $E_{\mathrm{J2}}$. An added benefit is that phase slips are dominant in the regime of equal Josephson energies, so that can be mitigated too by using unequal junctions. 

\end{document}